\documentclass{vgtc}                          

\graphicspath{{figures/}{pictures/}{images/}{./}} 

\usepackage{times}                     

\usepackage{tabu}                      
\usepackage{booktabs}                  
\usepackage{lipsum}                    
\usepackage{mwe}                       

\usepackage{mathptmx}                  
\usepackage{amsmath}
\usepackage{caption}
\usepackage{subcaption}
\onlineid{1058}

\vgtccategory{Research}

\vgtcinsertpkg

\title{ImageSI: Semantic Interaction for Deep Learning Image Projections}

\author{Jiayue Lin\thanks{e-mail: jiayuelin@vt.edu}\\ %
        \scriptsize Virginia Tech %
\and Rebecca Faust\thanks{e-mail: rfaust1@tulane.edu}\\ %
     \scriptsize Tulane University %
\and Chris North\thanks{e-mail: north@vt.edu}\\ %
     \scriptsize Virginia Tech}

\abstract{

    Semantic interaction (SI) in Dimension Reduction (DR) of images allows users to incorporate feedback through direct manipulation of the 2D positions of images. Through interaction, users specify a set of pairwise relationships that the DR should aim to capture. Existing methods for images incorporate feedback into the DR through feature weights on abstract embedding features.  However, if the original embedding features do not suitably capture the users' task then the DR cannot either. We propose, ImageSI, an SI method for image DR that incorporates user feedback directly into the image model to update the underlying embeddings, rather than weighting them. In doing so, ImageSI ensures that the embeddings suitably capture the features necessary for the task so that the DR can subsequently organize images using those features. We present two variations of ImageSI using different loss functions - ImageSI$_{\text{MDS$^{-1}$}}$, which prioritizes the explicit pairwise relationships from the interaction and ImageSI$_{\text{Triplet}}$, which prioritizes clustering, using the interaction to define groups of images. Finally, we present a usage scenario and a simulation-based evaluation to demonstrate the utility of ImageSI and compare it to current methods. 
     
} 


\newcommand{\rf}[1]{\textcolor{blue}{}}

\begin{document}


\firstsection{Introduction}
\maketitle

Sensemaking of image data is challenging due to the complex nature of images and the need to sequentially inspect images~\cite{gu2017visualization}. Dimension reductions (DR) help by identifying similarities and illustrating them with spatial proximity~\cite{cunningham2008dimension}. To enable DR for images, image embeddings must first be extracted using deep learning methods such as CNNs~\cite{lecun2015deep, 7780459}. However, DRs rely on accurate image embeddings; if the embeddings do not capture the image features well, the DR will not either. Thus, static DRs may not adequately support the users' tasks or reflect their prior knowledge.  

Semantic interaction (SI) describes a class of interaction methods that aim to infer the semantic meaning behind user interactions and adjust the visualization model according to user intents~\cite{10.1145/2207676.2207741}. In DR, SI enables users to directly interact with DR visualizations to convey feedback and update the DR model~\cite{10.1145/3158230}, to create a DR space that best reflects the users' tasks and knowledge.   
A recent approach by Han et al. enabled SI for DRs of images using a Weighted Multi-Dimensional Scaling (WMDS) approach~\cite{han2023explainable}. Their method applies weights to the data space before projection. During SI, the DR learns new projection weights that best capture user feedback and then applies them for re-projection. 
However, this relies on the embeddings adequately capturing the relevant features such that weighting them emphasizes the desired feature. In practice, this has limited ability to capture complex interactions and image features (see \cref{fig:merged_scatter_quant})~\cite{han2023explainable}.

To overcome this, we propose ImageSI, an SI framework for image DRs that fine-tunes the underlying embedding features to incorporate feedback, rather than weighting them.  Users still interact directly with the DR plot to specify feedback, but ImageSI now incorporates this feedback directly into the embeddings, ensuring that the embedding captures relevant features before projection.  By updating the embeddings rather than the DR, ImageSI discovers new or under-emphasized features in the data space that correlate to the users' feedback. This allows users to explore alternate embeddings that capture secondary features that are not well-represented by the original embeddings but are relevant to the users' task. Thus, ImageSI aims to create an embedding space that best matches the users' conceptual space, before DR, enabling the subsequent creation of a relevant DR space. 

ImageSI provides two different loss functions for incorporating feedback: MDS$^{-1}$, which aims to match the spatial organization defined by the interaction, and coordinate triplet margin loss, which emphasizes the creation of clusters based on user feedback. MDS$^{-1}$ better supports tasks on data that has more continuity, rather than discrete, disjoint classes. In contrast, coordinate triplet loss better supports tasks that rely on distinct groups of images. 

Our contributions are (1) the ImageSI framework for incorporating feedback directly into the image embeddings (2) two loss functions to support a wider range of tasks, (3) a usage scenario illustrating ImageSI, and (4) a quantitative comparison of ImageSI against current methods. 

\begin{figure*}[th]
    \centering
    \includegraphics[trim={0 10.5cm 0 5cm},clip, width=.9\linewidth]{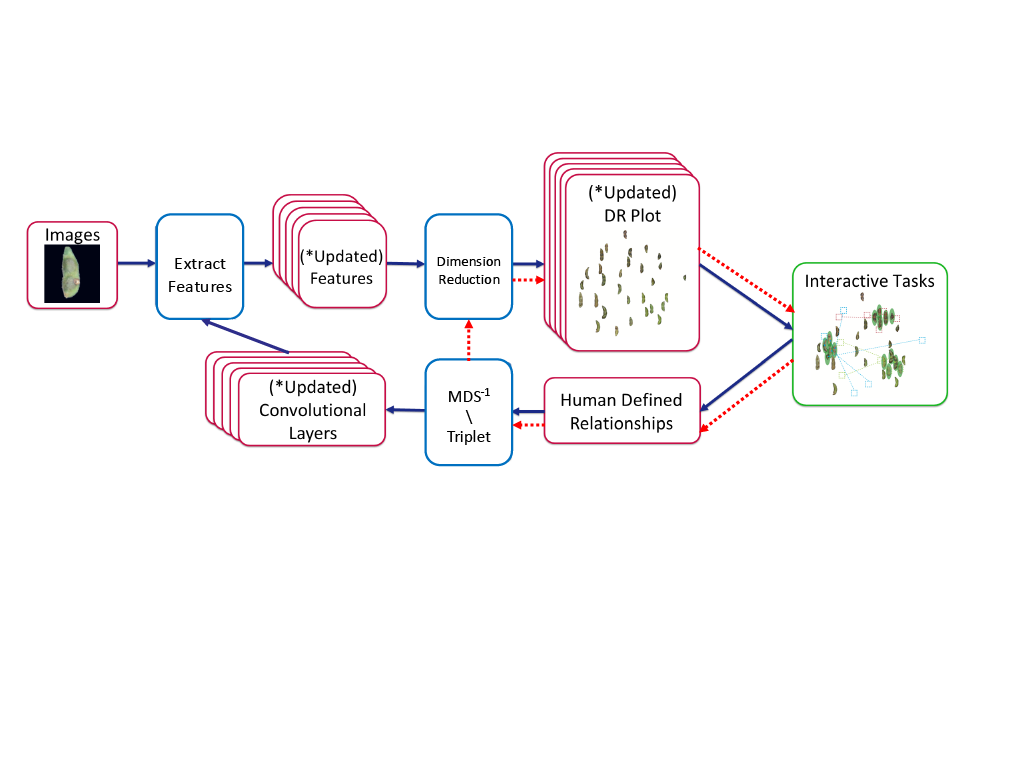}
    \vspace{-1em}
    \caption{The ImageSI pipeline. First, features are extracted from a pre-trained ResNet-18 model. These features are projected using DR. The user then performs interactive tasks on the DR plot. Their interaction is then used to fine-tune the image model using either MDS$^{-1}$ or triplet loss. Subsequently, ImageSI extracts the updated features from the fine-tuned model and re-projects them. Red dotted arrows represent the methodology from Han et al.~\cite{han2023explainable}, while blue solid arrows illustrate the ImageSI pipeline, which expands the scope of Han et al.'s exploration.} 
    \vspace{-1.5em}
\label{fig:ImageSI_pipeline_edited}
\end{figure*}


\section{Background and Related Work}
We discuss literature related to interactive DR and semantic interaction for visual analytics.

\textbf{Interactive DR }
Many methods exist for interactive DR~\cite{7536217}. We will briefly discuss the most relevant methods. Most similar to our approach are methods that enable steerable DRs. Some methods define and organize control points used to seed the DR~\cite{paulovich2011piece, 6065024, mamani2013user} while others directly learn a new distance function used by Multi-Dimensional Scaling (MDS)~\cite{10.1145/3158230,6400486}. 

\textbf{Semantic Interaction }
Semantic interaction (SI) aims to interpret the semantic meaning behind interactions with visualizations to adjust the visualization model based on user intents~\cite{10.1145/2207676.2207741}.  SI supports sensemaking by capturing the analytical reasoning of the user and applying it to their visual analysis~\cite{6327294}.  Andromeda enables SI in DRs using Weighted Multi-Dimensional Scaling (WMDS)~\cite{10.1145/3158230}. In applying weights to features, Andromeda allows people to change the importance of high-dimensional features in the 2D DR plot. It enables two forms of interaction: (1) directly changing the feature weights and (2) manipulating points in the DR plot to convey information and learning new weights via WMDS$^{-1}$.  This first method only works if the data has interpretable features.

\textbf{Semantic Interaction for Deep Learning DRs }
Recent work explores expanding on Andromeda to support SI in DRs of more complex data extracted from deep learning models, e.g. for text or images.  Han et al. presented a semantic interaction method for DRs of image data~\cite{han2023explainable}. Like Andromeda, their proposed pipeline uses WMDS to weight deep learning image features, enabling interaction through direct manipulation of points in the DR plot and learning updated weights through WMDS$^{-1}$. Additionally, they provide explanations of learned features through weighted saliency maps. The crux of this method is WMDS$^{-1}$ which learns new weights for WMDS based on the interaction such that the pairwise distances in the weighted feature space are proportional to the pairwise distances of the manipulated points. However, in practice, learning weights on abstract image embedding features has a limited capacity to capture complex human feedback~\cite{han2023explainable}.

Recent approaches enable SI in DRs of deep learning embeddings~\cite{10.1145/3397481.3450670,10.1145/3311790.3396646}. Most related, Bian et al.'s DeepSI enables SI to update the embedding features in deep learning DRs of text~\cite{10.1145/3397481.3450670}. Rather than learning DR weights, they fine-tune the underlying embeddings to better capture the semantics of the user feedback.
Our framework builds on these methods to enable semantic interaction in deep learning DRs of images that incorporate feedback directly into the image model, thus capturing more complex feedback. 

\section{ImageSI Framework}

To overcome the limitations of past methods, ImageSI proposes an SI method that incorporates feedback directly into the image model, rather than relying on re-weighting the feature space before projection. This approach has two primary advantages: (1) it captures more complex features that re-weighting static image features cannot and (2) it retains feedback between consecutive interactions, enabling incremental refinement to best incorporate the user's prior knowledge of the task. 

\subsection{User Workflow and Interaction}
\cref{fig:ImageSI_pipeline_edited} shows an overview of this workflow. First, ImageSI uses a deep learning image model to extract high-dimensional feature embeddings.  Then, it projects these features using a dimension reduction method to create a visual summary of the image collection and illustrates similarities via spatial proximity. Here, we use MDS~\cite{torgerson1958theory} but any DR could be substituted. 

To incorporate user knowledge or tune DRs to specific tasks, ImageSI enables users to directly organize points in the DR plot to specify relationships between sets of images. For example, in \cref{fig:initial_animal}(b), the user identifies an image feature, ``open-mouthed'' vs ``closed-mouthed'' animals, and conveys this feature by dragging together images of open-mouthed animals in one corner, and images of closed-mouthed animals in the other. ImageSI then incorporates this feedback into the underlying image model using one of the loss functions described below and re-projects the images to highlight this feature, shown in \cref{fig:animalmouth}.

\subsection{Image Model}
ImageSI uses pre-trained ResNet-18 as the underlying image model. While ImageSI uses ResNet-18, it generalizes to any image model.  Because the task is DR, not classification, it removes the last fully connected layer, leaving only the feature extraction layers \cite{8474912}. Based on user interactions, ImageSI fine-tunes the remaining layers to incorporate user feedback.


\subsection{Loss Functions}
ImageSI presents two variations,ImageSI$_{\text{MDS$^{-1}$}}$ and ImageSI$_{\text{Triplet}}$, using different loss functions to incorporate feedback.

\subsubsection{ImageSI$_{\text{MDS$^{-1}$}}$} 
Based on Bian et al.'s DeepSI~\cite{10.1145/3397481.3450670}, ImageSI$_{\text{MDS$^{-1}$}}$ defines a loss function such that, for every pair of points $y_i$ and $y_j$ organized by the interaction, the pairwise distances in the embedding space closely match the pairwise distances specified in the 2D DR space. It does this by learning model weights, $w$, such that for an image $x_i$, $M(x_i,w)$ gives the image embedding for $x_i$, using the model $M$.  The updated weights are given by: 
\vspace{-.5em}
\begin{equation}
\underset{w}{\arg \min} \sum_{i < j \leq N} \left( \text{dist}_{L}(y_i, y_j) - \text{dist}_{H}(M(x_i,w), M(x_j,w)) \right)^2
\vspace{-.5em}
\end{equation} 

where $dist_L(y_i,y_j)$ is the 2D distance from the interaction, and $dist_H(x_i,x_j)$ is the HD distance between the image embeddings. We call this loss function MDS$^{-1}$ as it uses the same stress function as MDS to optimize the model weights.

MDS$^{-1}$ supports tasks where the user-defined organization contains meaningful order and spatial relationships, in contrast to discrete image sorting tasks. However, MDS$^{-1}$ is sensitive to the configuration of the manipulated points in the DR space and thus has more difficulty picking up on secondary structures not explicitly defined by the interaction. 

\subsubsection{ImageSI$_{\text{Triplet}}$}
For less constrained tasks that aim to organize images into broader groups, rather than an explicit spatial ordering, ImageSI$_{\text{Triplet}}$ uses Coordinate Triplet Margin Loss. Traditional triplet margin loss \cite{7298682, hoffer2015deep} relies on labeled classes to identify triplets of images containing an anchor image, a positive image (of the same class), and a negative image (of a different class), see ~\cite{7298682} for detail. However, exploratory sensemaking tasks may not operate on labeled data and thus, we do not want to assume that users have labeled data. 

To overcome this, we designed Coordinate Triplet Margin Loss which infers positive and negative samples based on the distances between the 2D coordinates of the points organized during interaction. ImageSI forms triplets from the points moved during interaction ($V$), using each point $a\in V$ as an anchor. 
Triplets of images are chosen such that, for a given anchor $a$, ImageSI first generates a pool of positive ($P$) and negative ($N$) samples by calculating the absolute differences between the coordinates of $a$ and the coordinates of all other points in $V$ and then randomly chooses a positive and negative sample from these pools. 
The positive pool $P$ is generated by $P = \{v | v\in V-a \text{ where } d(a,v) < \epsilon_p\}$ where $d(a,v)$ is the Euclidean distance between the anchor ($a$) and the other point ($v$) and $\epsilon_p$ is a threshold for the maximum distance allowed to be considered a positive (similar) sample. 

The negative pool $N$ is generated similarly, with $N = \{v | v\in V-a \text{ where } d(a,v) > \epsilon_n\}$ where $\epsilon_n$ is a threshold for the minimum distance needed to be considered a negative sample. This process generates a pool of positive and negative samples to form the triplets. 

\section{Usage Scenario}
\label{sec:case}
In this section, we demonstrate the practical applications of ImageSI by comparing ImageSI$_{\text{MDS$^{-1}$}}$ and ImageSI$_{\text{Triplet}}$, against the baseline model from Han et al., which we will call WMDS$^{-1}$. We evaluate their performance in a real-world scenario where the initial DR fails to capture a desired feature of the images. 

\begin{figure}
    \centering
    \begin{subfigure}[b]{.49\linewidth}
        \centering\includegraphics[height=\linewidth,width=\linewidth,keepaspectratio]{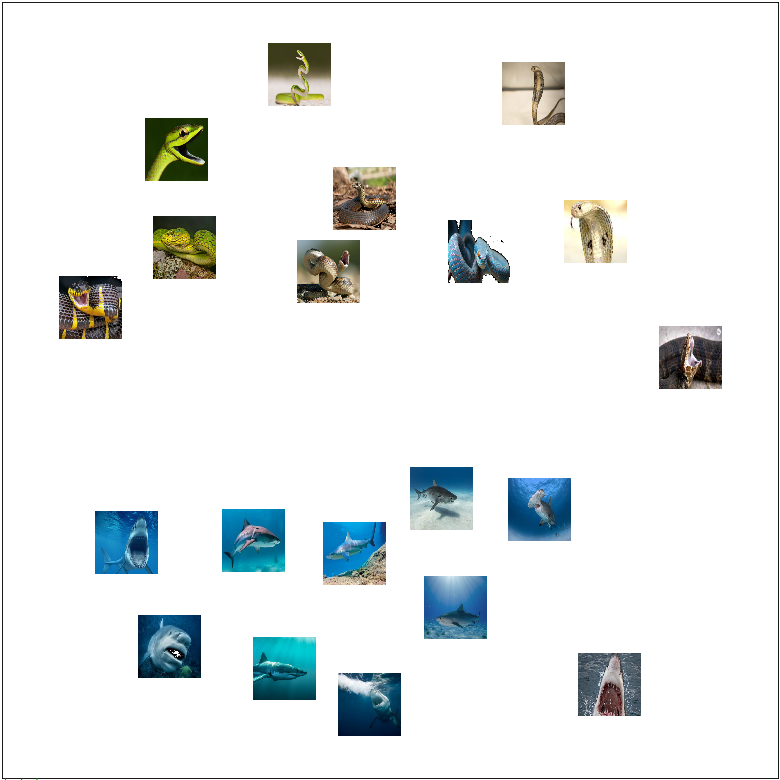}
        \caption{Initial Projection}
    \end{subfigure}
    \begin{subfigure}[b]{.49\linewidth}
        \centering
        \includegraphics[height=\linewidth,width=\linewidth,keepaspectratio]{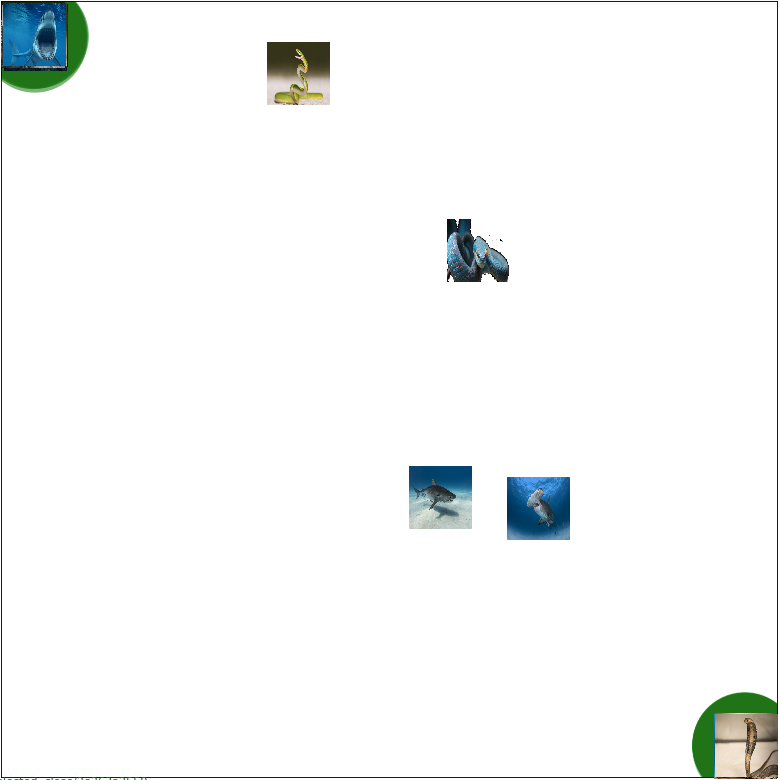} 
        \caption{User Manipulation}
    \end{subfigure}
    \vspace{-.5em}

    \caption{(a) The initial MDS projection of the images containing open and closed-mouthed sharks and snakes. (b) The semantic interaction teaches the DR about the open vs closed mouth feature.}
    \vspace{-1.5em}
    \label{fig:initial_animal}
\end{figure}

\begin{figure*}[t]
\vspace{-.8em}
    \centering
    \begin{subfigure}[b]{0.3\linewidth}
        \centering
        \includegraphics[trim={0 .5cm 0 .5cm},clip,height=\linewidth,width=\linewidth,keepaspectratio]{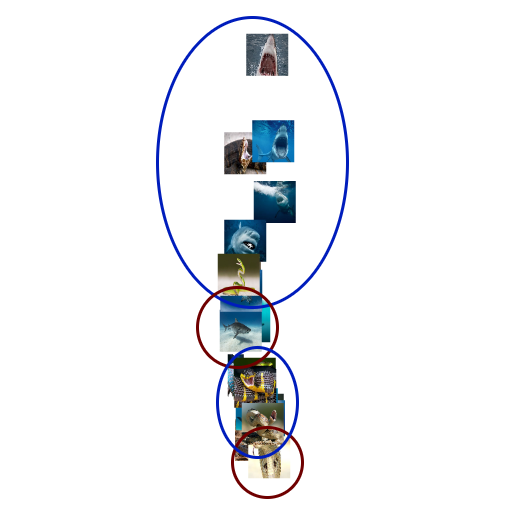}
        \caption{WMDS$^{-1}$}
    \end{subfigure}
    \begin{subfigure}[b]{0.3\linewidth}
        \centering
        \includegraphics[trim={0 .5cm 0 .8cm},clip,height=\linewidth,width=\linewidth,keepaspectratio]{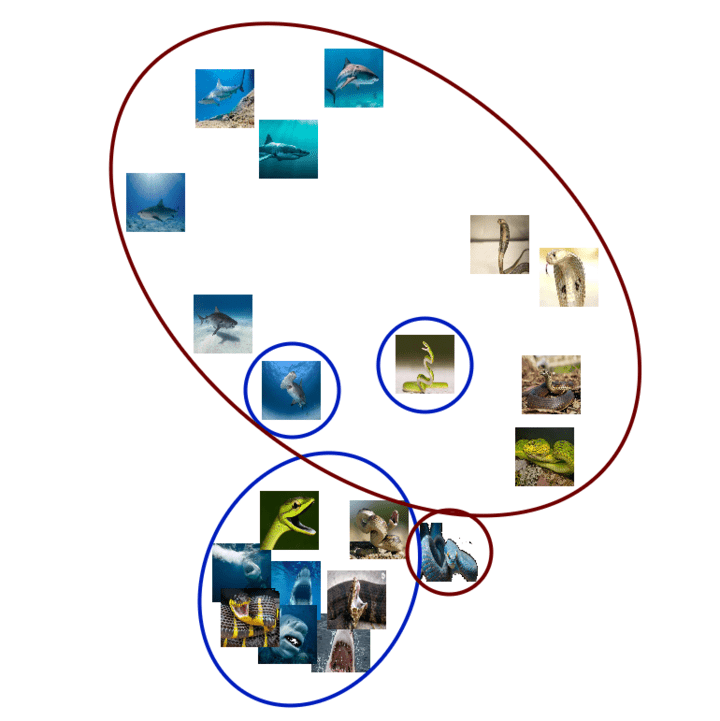} 
        \caption{ImageSI$_{\text{MDS$^{-1}$}}$}
    \end{subfigure}
    \begin{subfigure}[b]{0.3\linewidth}
        \centering
        \includegraphics[trim={0 .5cm 0 .4cm},clip,height=\linewidth,width=\linewidth,keepaspectratio]{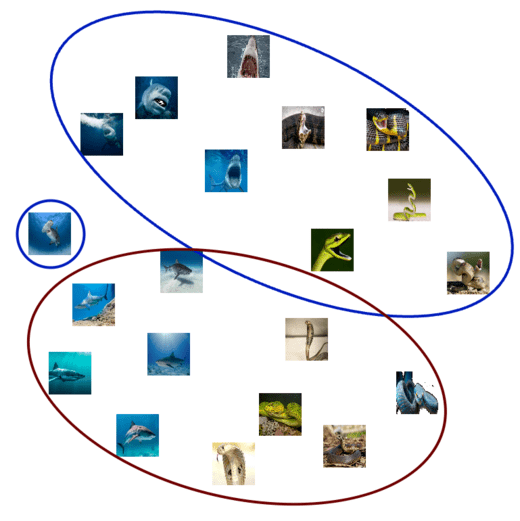}
        \caption{ImageSI$_{\text{Triplet}}$}
    \end{subfigure}
    \vspace{-.5em}
\caption{Updated DR plots after interaction for (a) WMDS$^{-1}$,  (b) ImageSI$_{\text{MDS$^{-1}$}}$, and  (c) ImageSI$_{\text{Triplet}}$. Note, blue ellipses indicate the open-mouth animals, while red indicates closed-mouth animals.}
\vspace{-1.5em}
\label{fig:animalmouth}
\end{figure*}

\textbf{Data: } We start by loading a dataset comprising images of sharks and snakes, shown in \cref{fig:initial_animal}(a). Initially, the DR organizes the images based on the animal type, a snake or shark. Upon inspection, we notice that some animals have open mouths while others have closed mouths. We want to re-organize the images to capture this feature, rather than organizing them by animal type.

\textbf{Interaction: } To convey this information to the model, we perform organize several images of each type into distinct regions. As shown in \cref{fig:initial_animal}(b), we select 8 animals with open mouths (4 sharks, 4 snakes) and 8 animals with closed mouths (4 sharks, 4 snakes), positioning them in opposing corners to convey their distinctions. \cref{fig:animalmouth} shows the updated DR under all three models. 

\textbf{Results: } \cref{fig:animalmouth}(a) shows the updated projection for the baseline method, WMDS$^{-1}$. Blue contours enclose animals with open mouths, while red contours enclose animals with closed mouths. 
We see that WMDS$^{-1}$ failed to create a clean separation of the two features, open-mouthed and closed-mouthed. It creates a nearly linear organization of the images where most of the sharks are spread in the top half and most of the snakes in the bottom. It partially separates by open vs closed within each animal type, the open-mouthed sharks are spread along the top portion but fail to organize the snakes in a meaningful way.

\cref{fig:animalmouth}(b), shows the updated projection for ImageSI$_{\text{MDS$^{-1}$}}$. We see significant improvement over the previous method, creating two more distinct groups of images using the specified feature. Open-mouth animals are predominantly grouped at the bottom, while closed-mouth animals cluster in the middle right and top portions. However, it mis-projects a few images along the boundary between the two features.

\cref{fig:animalmouth}(c) shows the updated projection for ImageSI$_{\text{Triplet}}$. Now we see a distinct grouping based on the specified feature. The images are well separated by the ``open-mouthed'' feature, but still spread throughout the DR space to create a readable plot. Additionally, this method not only separated the images using the specified feature but also organized the images within the group by a secondary feature, their animal type. For this task, ImageSI$_{\text{Triplet}}$ best captured the user feedback while also creating a secondary organization by other identified features (i.e., animal type). 

\section{Quantitative Simulation-Based Evaluation}
To quantitatively compare the performance of the three models: ImageSI$_{\text{MDS$^{-1}$}}$, ImageSI$_{\text{Triplet}}$, and WMDS$^{-1}$, we conducted a simulation-based evaluation to measure how well each method captures human feedback. We focus on the task of separating images by distinct visual features, evaluating how well the updated DR clusters are based on those features. We also examine how the number of interactions affects the quality of the resulting DR. 




\subsection{Method}
Following prior research \cite{han2023explainable}, we create a simulation engine that simulates SI. It performs interactions to arrange images such that those belonging to the same category are grouped close together, while those from different categories are distinctly separated.  Using this interaction, it updates the model (or the DR weights in the baseline case) using the specified SI method and re-projects the images. We then evaluate the clusters created by the interaction against their ground truth labels, using a clustering metric. We repeat this simulation many times, varying the number of points used in the interaction. Note, 
for ImageSI$_{\text{Triplet}}$, we need at least two samples per category in the dataset to ensure anchor positive pairs. 



\subsection{Simulation Engine}

To evaluate the performance of the system and assess the effectiveness of semantic interactions, we employ a simulation engine comprising two key components: the interaction simulator and the projection evaluator.

{\textbf{Interaction Simulator:}}
The interaction simulator simulates semantic interactions to guide the layout of image datasets. For ImageSI$_{\text{MDS$^{-1}$}}$ and WMDS$^{-1}$, the simulator simply selects the specified number of images, $k$, in each class (``open-mouthed'' or ``closed-mouthed'') and generates a distance matrix such that, for two points $x_i$ and $x_j$, $|| x_i - x_j ||$ is 0 if the $x_i$ and $x_j$ are from the same class and $\sqrt{2}$ otherwise. 

For ImageSI$_{\text{Triplet}}$, it selects $k$ samples from each class. It then randomly picks an anchor sample, though each selected sample will be used as an anchor in turn. A positive sample, different from the anchor point, is randomly chosen from the same class as the anchor. Similarly, a negative sample is randomly selected from the points in the other classes. This requires that at least two points per class are moved ($k>=2$), as selecting only one point would not provide a positive sample for the randomly selected anchor points.

After simulating the interaction, we apply the corresponding loss function to fine-tune the model. Finally, we extract the updated image embeddings and re-project them using MDS. 



{\textbf{Projection Evaluator:}}
After re-projecting the new embeddings, the layout evaluator assesses the quality of the projected layout. To evaluate the effectiveness of ImageSI in capturing simulated user feedback, we utilize an adjusted Silhouette score \cite{rousseeuw1987silhouettes}. 
The Silhouette score evaluates clustering quality by considering the tightness of points within clusters (cohesiveness) and the distinctiveness between clusters (separation). The Silhouette score spans from -1 to 1. Near-zero scores suggest cluster overlap, negatives imply mis-assignments, and positives indicate well-separated clusters \cite{LEFRANC201611}.

Following Han et al.'s approach \cite{han2023explainable}, we use an adjusted silhouette score that better suits sensemaking tasks. Similar to them, our aim is not to create tight, well-separated clusters as valuable information may be contained in the spread of clusters and we do not assume that the projected data contains distinct, disjoint clusters. Thus, our target Silhouette score is around 0.5, preferring arrangements where data points are typically twice as far from the closest class as they are from their own class. To emphasize this preference, we adjust the Silhouette score by doubling it so that an ideal score is one. Scores between zero and one signal too much spreading, while scores above one suggest excessive clustering.




A higher adjusted Silhouette score signifies better clustering performance, with values nearing 1 suggesting well-separated clusters, while those nearing 0 imply overlapping clusters. Scores near -1 imply potential mis-assignments.




\subsection{Dataset and Task}
We expand the dataset from the case study in \cref{sec:case} to consist of 40 images, with 10 open-mouth sharks, 10 open-mouth snakes, 10 closed-mouth sharks, and 10 closed-mouth snakes. The tasks remain the same, organizing animal images based on open and closed mouths. We run the simulation engine 10 times for each model and average the adjusted Silhouette score to obtain a final robust result.

\begin{figure}[t]
    \centering

    \includegraphics[trim={0 0 0 1cm},clip,width=\linewidth]{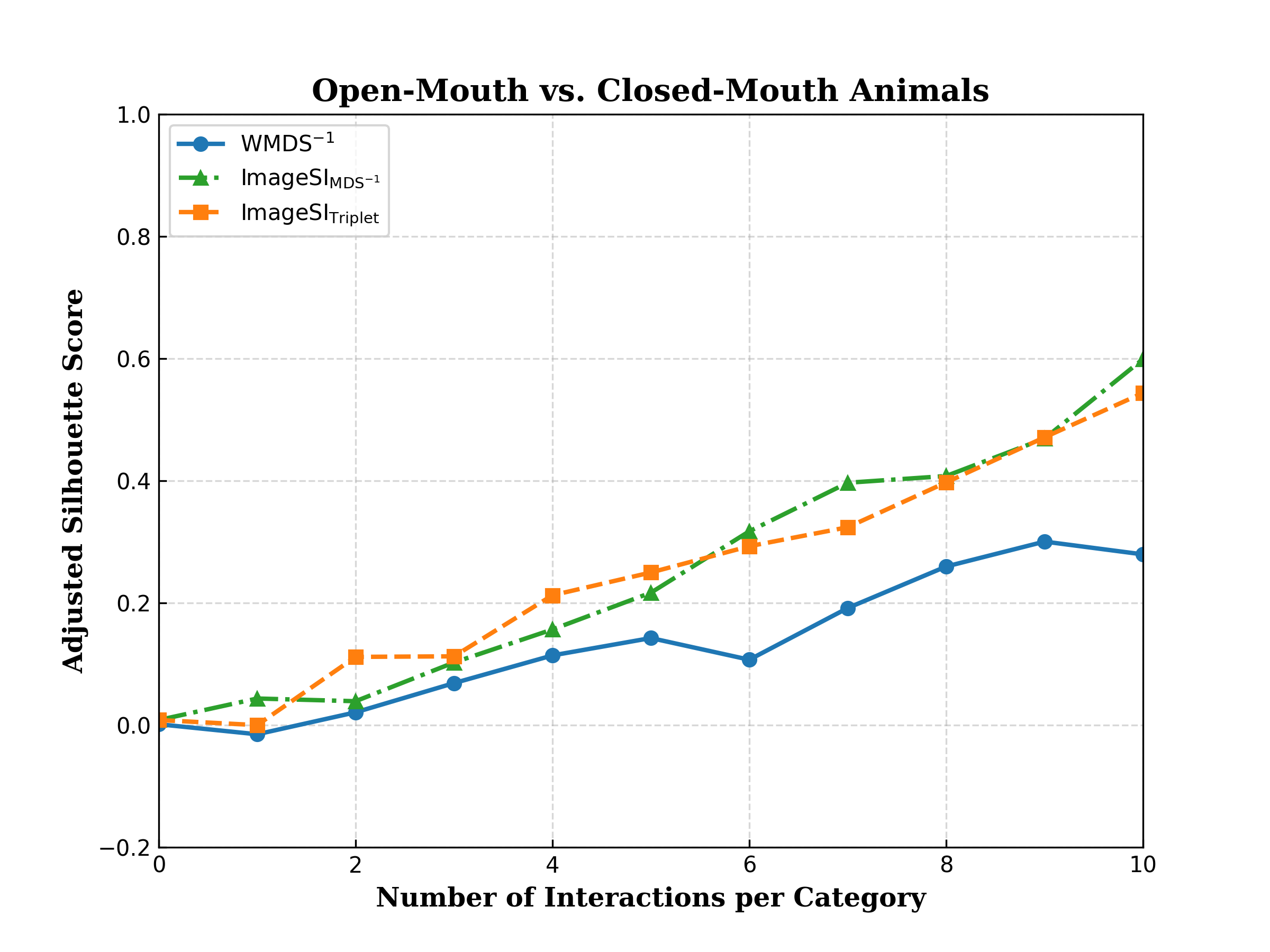}
    \vspace{-2.5em}
    \caption{Comparison of adjusted Silhouette scores across different frameworks and tasks. Subfigures (a) to (d) depict the performance of WMDS$^{-1}$, ImageSI$_{\text{MDS}^{-1}}$, and ImageSI$_{\text{Triplet}}$, respectively. Each subplot shows the adjusted Silhouette scores achieved by each method over a range of interactions.}
      \vspace{-2em}
\label{fig:merged_scatter_quant}
\end{figure}

\subsection{Result}
\cref{fig:merged_scatter_quant} shows the adjusted Silhouette scores for varying numbers of interactions across the three different models. We see that WMDS$^{-1}$ struggles to get above a score of $\approx 0.3$. This is likely caused by the overlapping behavior we saw in \cref{fig:animalmouth}(a).  In contrast, both ImageSI$_{\text{Triplet}}$ and ImageSI$_{\text{MDS}^{-1}}$ outperform mageSI$_{\text{WMDS\(^{-1}\)}}$, continually improving their layout with each interaction. For reference, the plot in \cref{fig:animalmouth} (c) has a score of 0.653, similar to the final scores reached by ImageSI$_{\text{Triplet}}$ and ImageSI$_{\text{MDS}^{-1}}$. This validates that incorporating the feedback directly into the embeddings better captures the intent of user interactions over prior methods and produces DRs more relevant to the user's task.

\section{Discussion}

\textbf{Choice of Loss Function: }In interactive deep metric learning, a traditional metric loss like triplet loss \cite{7298682}, contrastive loss \cite{1640964}, angular loss \cite{wang2017deep}, quadruplet loss \cite{chen2017beyond}, N-Pair loss \cite{NIPS2016_6b180037}, and Histogram loss \cite{NIPS2016_325995af} have been used to shape the representation learned by the model. In this work, we employ triplet loss and MDS$^{-1}$ to guide the DL model in capturing user intention. Triplet loss optimizes the embedding space based on relative distances between samples. However, it largely disregards the actual pairwise distances between data points, only using them to infer clusters of images, which potentially overlooks meaningful feedback. In contrast, MDS$^{-1}$ aligns pairwise distances in the embedding space with those in the DR space but does not create as cleanly organized groups of images. Integrating MDS$^{-1}$ with triplet loss would address this limitation by incorporating pairwise distances into the learning process, while still placing an emphasis on clustering. This would pair the detailed feedback from MDS$^{-1}$ with the cluster's superior organizational abilities of the triplet. The integration involves using recovered pairwise distances to guide learning, enhancing the model's ability to effectively capture local and global structures.

\textbf{Tradeoffs between WMDS$^{-1}$ and ImageSI: }
While ImageSI showed superior performance to the baseline WMDS$^{-1}$, there is a trade-off between these two methods. Because ImageSI incorporates feedback directly into the model, it retains previous feedback, allowing users to iteratively tune the embeddings. While useful when performing incremental, related interactions, interactions that are unrelated or contradictory to prior ones may confuse the model and result in a worse embedding. In contrast, WMDS$^{-1}$ supports isolated, rapid adjustments for different tasks, without needing to reset the model in between. However, it does not support incremental refinement of the DR and less effectively captures feedback.

\textbf{ Conclusion and Future Work: }
In this paper, we presented ImageSI, a framework for the SI of image DRs that incorporates feedback directly into the image embeddings. We showed ImageSI's superior performance to past methods at incorporating feedback into the DR pipeline. In future work, we will explore the creation of custom loss functions to better incorporate feedback. Additionally, we will investigate methods to introduce explainability to validate the information learned by the interaction. These improvements will enable a more effective, human-in-the-loop DR for image sensemaking. 

\section*{Supplemental Materials}
\label{sec:supplemental_materials}
All supplemental materials are available on OSF at \url{https://osf.io/m2wdf/?view_only=3b2f851592874ac791ad0ba5bc809774}, released under a CC BY 4.0 license.  This includes a document with three additional usage scenarios. .

\acknowledgments{
This material is based upon work supported by the National Science Foundation under Grant \# 2127309 to the Computing Research Association for the CIFellows 2021 Project.}

\bibliographystyle{abbrv-doi-hyperref}

\bibliography{template}
\end{document}